\documentclass[useAMS,usenatbib]{mn2e}
\usepackage{graphics,graphicx,epsfig,psfig}
\usepackage[normalem]{ulem}
\usepackage[dvipsnames]{color}
\usepackage{soul,xcolor}
\usepackage{amsmath, amssymb}
\usepackage[]{inputenc,amssymb}
\usepackage{booktabs,caption}
\usepackage{amsmath}
\usepackage{multirow}
\usepackage{soul}
\usepackage{subfig}

\setstcolor{red}

\def \be{\begin{equation}}
\def \ee{\end{equation}}
\def \ba{\begin{eqnarray}}
\def \ea{\end{eqnarray}}
\def \etal{{et al.}}

\definecolor{webgreen}{rgb}{0,.5,0}
\definecolor{webbrown}{rgb}{.6,0,0}

\usepackage[%
    colorlinks = true,%
    linkcolor = blue,%
    urlcolor  = blue,%
    citecolor = webgreen,%
    anchorcolor = blue]{hyperref}

\newcommand{\ufhref}[3][blue]{\href{#2}{\color{#1}{#3}}}%

\title[Superbubble size distribution]{Size distribution of superbubbles}
\author[B. B. Nath, P. Das, M. S. Oey]
{Biman B. Nath$^1$, Pushpita Das$^{1,2}$, M. S. Oey$^{3}$\\
$^1$ Raman Research Institute, Sadashiva Nagar, Bangalore 560080, India\\
$^2$ Indian Institute of Science Education \& Research Kolkata, Mohanpur 741246, India\\
$^3$ Dept of Astronomy, University of Michigan, 323 West Hall, 1085 South University Ave., Ann Arbor, MI  48109-1107, USA\\
}

\begin{document}


\maketitle

\label{firstpage}

\begin{abstract}
We consider the size distribution of superbubbles in a star forming galaxy. Previous studies have tried to explain the
distribution by using adiabatic self-similar evolution of wind driven bubbles, assuming that bubbles stall when pressure
equilibrium is reached. We show, with the help of hydrodynamical numerical simulations, that this assumption is not valid.
We also include radiative cooling of shells. In order to take into account non-thermal pressure in the ambient medium, we
assume an equivalent higher temperature than implied by thermal pressure alone. Assuming that bubbles stall when the 
outer shock speed becomes comparable to the ambient sound speed (which includes non-thermal components), we recover the size distribution with a slope of $\sim -2.7$ for typical values of ISM pressure in Milky Way, which  is consistent with observations.
Our simulations also allow us to follow the evolution of size distribution
in the case of different values of non-thermal pressure, and we show that the size distribution steepens with lower pressure, 
to slopes intermediate between only-growing and only-stalled cases.
\end{abstract}


\section{Introduction}
The distribution of sizes of superbubbles created by stellar winds and supernovae of OB associations can be a 
diagnostic of the star formation process in a galaxy. The seminal paper by \cite{oey1997} [hereafter OC97] showed that if the mechanical luminosity
function of OB associations is described by a power law , $\phi(L) \propto L^{-\beta}$, then for a constant star formation rate,
the differential size distribution of superbubbles is given by $N(R) \propto R^{1-2\beta}$, for bubbles whose evolution is dominated by
ambient pressure and have stalled. For typical parameters, OC97 estimated this stalling radius to be $ \le 1$ kpc, which implies that
observed superbubbles are in this phase of evolution. This robust prediction of the size distribution to have a power-law with index $1-2\beta$ 
makes it a useful diagnostic of the star formation process in a galaxy. This predicted distribution has been confirmed by {\it The HI Nearby Galaxy Survey}, which studied the properties of HI holes in nearby galaxies \citep{bagetakos2011}. They found that the size distribution has a power-law index of $\sim-2.9$, which implies $\beta\approx 2$, which in turn is consistent with independent observations of OB association \citep{mckee1997}.

The robustness of this predicted size distribution, however, raises the question whether or not it depends on parameters such as ambient pressure, 
density and so on, and if so, how. Since it was derived assuming that ambient pressure does not affect the adiabatic expansion of a stellar wind bubble \citep{weaver1977} other than to impose a stall criterion, it is not possible to answer these questions without going beyond these assumptions. Although the average distribution
in {\it THINGS} came out to have a power law of $\sim -2.9$, individual galaxies had distributions whose power-law ranged between $-2$ and $-4$.
Early-type spiral galaxies showed a steeper slope than late-type spirals and dwarf galaxies. \cite{bagetakos2011} explained this by invoking the fact that scale heights of disks in early-type spirals are smaller than in late-type spirals, and are easy to break out of. This would limit the size of the largest holes, and consequently steepen the size distribution. However, one could also envisage a scenario in which a truncated or steepened OB association luminosity function in early type galaxies produce a steep size distribution of HI holes, as is known to also manifest in the HII region luminosity
  function \citep{oey1998}. In order to disentangle the effects, one would have to calculate the size distribution beyond the assumptions of adiabaticity, which is inherent in the self-similar evolution of bubbles. And then the question remains as to how the power-law distribution with index $\sim -3$ derived from such assumptions match the observations.

In this paper we relax the assumptions of self-similarity in the bubble evolution, include radiative cooling, with the help of 1-D hydrodynamic numerical simulation, and derive the size distribution. We assume a constant star formation rate for simplicity. 
The paper is structured as follows. We first review the results of OC97 in light of Monte-Carlo simulations in \S 2. Then we discuss the results when ambient pressure is included in the dynamics of superbubbles, in \S 3. In \S 4, we discuss the effects of cooling and present the results in \S 5. The implications are discussed in \S 6.

\section{Self-similar superbubbles}
We first review the essence of OC97 calculations. They assumed that the mechanical luminosity function of OB association is given by 
$\phi(L) \propto L^{-\beta}$. The luminosity of each cluster is assumed to be constant in time, until the lowest mass SN progenitors expire at $
t_e\sim 40$ Myr. The wind bubble is assumed to evolve in a self-similar manner, so that the radius scales as, $R \propto L^{1/5} t^{3/5}$
\citep{weaver1977}. Note that this assumes adiabatic expansion of the bubble. The
pressure inside then evolves as $p_i \propto L^{2/5} t^{-4/5}$. If the ambient pressure is $p_0$, then these bubbles are assumed to stall when $p_0=p_i$. This implies a stalling time, for the bubble to reach a final radius, of $t_f \propto L^{1/2}$. This in turn leads to a scaling of the final stalling radius $R_f \propto L^{1/5} t_f^{3/5} \propto L^{1/2}$. 

If the superbubbles are produced at a constant rate $\psi$, then after time $t$, the number of bubbles with radii in the range $R$ to $R+dR$ will be dependent on $\phi(L)dL$, where $L$ and $dL$ are the luminosity and range corresponding to this interval in radius $R$. Therefore the
differential size distribution will be given by,
\be
N(R) \propto \psi \phi(L) \Bigl ( {\partial R_f \over \partial L} \Bigr )^{-1} \, \propto L^{-\beta+1/2} \propto R^{1-2\beta}\,,
\label{weaver_distribution}
\ee
where the last proportionality follows from $R_f \propto L^{1/2}$ for stalled final radii derived above. 
Therefore the above scalings imply a size distribution $N(R) \propto R^{1-2\beta}$, for stalled bubbles. 
In case of expanding bubbles in the momentum conserving phase, one again has $R \propto L^{1/4} t^{1/2}$. In this case, OC97 considered 
stalling of bubbles when bubble speed becomes comparable to ambient sound speed (or, equivalently, ambient pressure being comparable
to ram pressure of the outer shock). This again leads to $t_f \propto L^{1/2}$, and $R_f \propto L^{1/2}$, leading to $N(R) \propto R^{1-2\beta}$. 
They showed that stalled bubbles dominate the size distributions.
They also estimated that for typical ISM parameters, the largest size of stalled bubbles is $\sim 1$ kpc, given the lifetime of OB associations. Bubbles larger than this would break out of the disc (the radius being much larger than the scale height), and need not be considered. Therefore the size distribution for typical ISM parameters would be dominated by stalled bubbles.

We show this size distribution with the help of  Monte-Carlo calculation. For simplicity we assume a uniform star formation rate of $1$ M$_\odot$ yr$^{-1}$. The luminosity function of OB association is taken as $\phi(L)=A L^{-\beta}$, with $\beta=2$, as inferred from HII region luminosity function \citep{oey1998, mckee1997}. The mechanical luminosity is related to the mass
of the cluster in the following way. Given a Kroupa initial mass function \citep{kroupa2002}, there is one OB star for $100$ M$_\odot$ cluster mass, and each
OB star can be assumed to give $10^{51}$ erg. The average luminosity of the cluster (given a lifetime of $\sim 37$ Myr, corresponding to the 
main sequence life time of a 8 M$_\odot$ star), is
\be
L\approx 9 \times 10^{33} \, {\rm erg} \, {\rm s}^{-1} \, \Bigl ( { M_{cl} \over M_\odot} \Bigr ) \,.
\ee
This is very close (within a factor of a few) to the result of mechanical luminosity of star clusters including the effect of stellar winds and supernovae, as calculated by \cite{leitherer1999}.
Therefore the mass function of the clusters can be written as 
\be
 {dN \over dM_{cl}}=A \Bigl ( {M_{cl} \over M_\odot} \Bigr )^{-\beta}\,.
 \ee
 Assuming the minimum and maximum mass of clusters to be $100$ and $10^6$ M$_\odot$ respectively, the average cluster mass is found
 to be $\sim 1360$ M$_\odot$, corresponding to an average mechanical luminosity of $\sim 1.2 \times 10^{37}$ erg s$^{-1}$. The luminosity range is $\sim 9 \times 10^{35}\hbox{--} 9 \times 10^{39}$ erg s$^{-1}$. 
 
\begin{figure}
\centering
\includegraphics[width=8.78cm, angle=0]{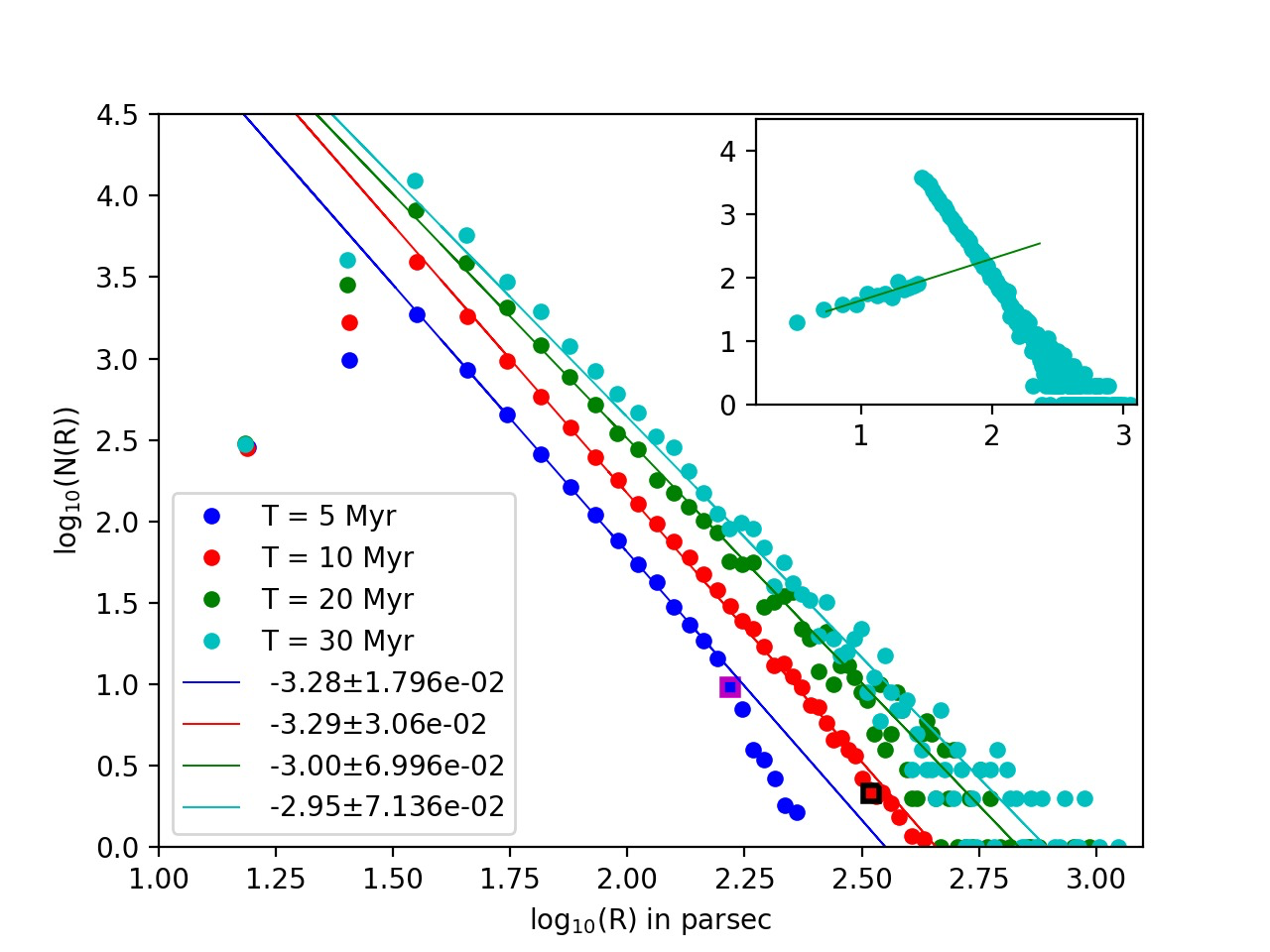}
\caption{Size distribution of self-similar superbubbles  for a constant star formation rate at different epochs, for an ambient pressure $p_0=2.76 \times 10^{-12}$ dynes cm$^{-2}$, and ambient temperature $T_{\rm amb}=4 \times 10^4$ K.  The square boxes on the blue and red lines show the values of the maximum stall radius at $5$ and $10$ Myr. The distribution at each epoch has been fitted with a power-law below the maximum size shown by the square boxes, and the fitted slope are all roughly close to $-3$, below the maximum stall radius. The inset shows the distribution at 30 Myr in detail (with smaller bin size), which has the $R^{2/3}$ rising part for small and growing bubbles (the straight line shows a power-law of index $2/3$).
}
\label{weaver}
\end{figure}

 The size distribution of bubbles is shown in Figure \ref{weaver}, for different times after the onset of star formation, for bubbles growing in a  self-similar manner as mentioned above. The ambient pressure is assumed to be $P_0=2.76 \times 10^{-12}$ dynes cm$^{-2}$ as was considered by OC97. As expected from the arguments in OC97, the dominant slope of the distribution is roughly $1-2\beta=-3$, for radii below the maximum size of stalled bubbles at a given epoch. 
OC97 estimated the stalling timescale and the stalled sizes. Initially, low luminosity bubbles stall, and at a given epoch, bubbles up to a certain size stall, beyond which the bubbles are in a growing phase. We show this upper limit of stalled sizes with square boxes on the red and blue lines, corresponding to $5$ and $10$ Myr (using equation 31 of OC97). (At later times, the biggest stalled bubbles cross the limit of sizes considered by us.) Figure \ref{weaver} shows that the fitted power-laws until this maximum size has an 
 index close to $-3$. However, there is also another class of small bubbles that are growing at any given epoch, arising from either massive but young clusters or low mass clusters.

We assume the bubbles to start forming at a continuous rate at $t=0=t_l$.
The inset in Figure \ref{weaver} plots the size distribution at 30 Myr with smaller bin size, to clearly show the case of small and growing bubbles. These bubbles show a distribution that scales as $R^{2/3}$ as predicted by OC97 (their eqn 39). These bubbles are dominated by clusters which have not yet reached stalling phase. In this case, $R \propto L^{1/5} t^{3/5}$, which gives ${\partial R \over \partial L} \propto L^{-4/5} t^{3/5}$.  This distribution is then integrated from $t_l$ until a time, $t_u$, for which the minimum luminosity cluster is still growing. This time scale is given by $t_u \propto R^{5/3} L_{\rm min}^{-1/3}$ (from self-similar evolution). 
Therefore, the size distribution of growing bubble is given by 
\be
N(R)  \propto \int^{t_u} L^{-\beta}\, ({\partial R \over \partial L})^{-1}  \,dt 
\propto R^{4-5\beta} t_u^{-2+3\beta} \propto R^{2/3}
\ee
Our Monte-Carlo runs for self-similar case show these growing bubbles with a size distribution as predicted.

The increasing ($R^{2/3}$) and decreasing $R^{-3}$) regimes of the size distribution  give rise to a peak in the size distribution, and the peak shifts towards large radii with time, until the time when  the oldest population of minimum luminosity clusters has achieved stalling radius. In the case depicted in Figure \ref{weaver} the minimum luminosity cluster achieves this status at $\sim 1$ Myr (using eqns 24 and 31 of OC97 for the minimum luminosity), after which the peak does not shift. The bubble size corresponding to the peak of the distribution is not small or negligible from an observational point of view. For example, after $5$ Myr, the peak is found to be at $\sim 40$ pc.

We note here that the assumption of constant luminosity with time for low mass clusters is justified because the mechanical luminosity not only
arises from supernovae, which in the case of low mass clusters would be few and far between, but also from stellar winds of massive stars. Since the typical mechanical power of stellar winds from OB stars is $\sim 10^{36}$ erg $s^{-1}$\citep{seo2018}, which is also coincidentally comparable to $(10^{51} \, {\rm erg}/35 \, {\rm Myr})$, the mechanical power of OB associations before and after the first supernovae events differ at the most by a factor of 2. This is also borne out by the estimates of mechanical power using Starburst99 \citep{leitherer1999}.

\section{Beyond zero-pressure ambient medium}
When we consider the growth of bubbles beyond the self-similar evolution, we need to fully account for the effect of ambient pressure and radiative
cooling. The effect of radiative cooling cannot be included without resorting to hydrodynamical simulation. However, the effect of ambient pressure can be calculated numerically, and before going to the full solution of hydrodynamical simulation, we will discuss this effect next.

\begin{figure}
   \centering
    \subfloat{{\includegraphics[width=4.5cm]{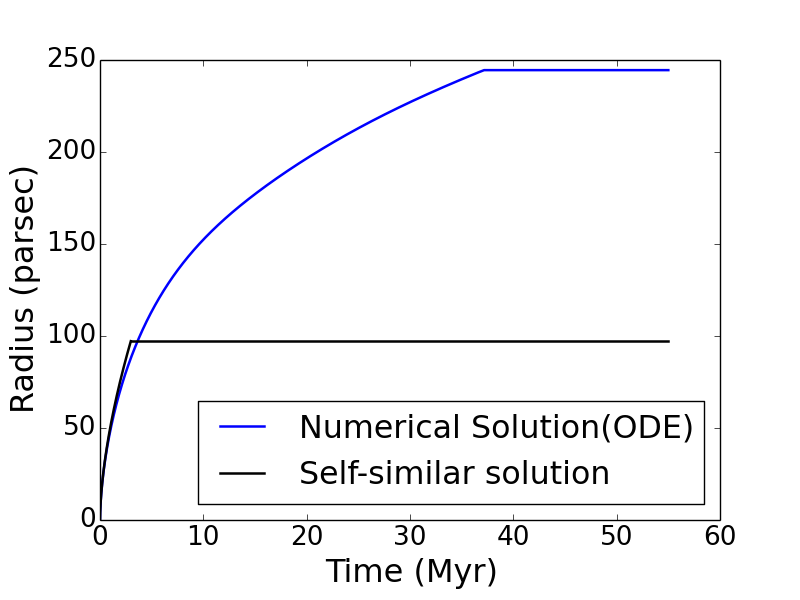} }}%
    \subfloat{{\includegraphics[width=4.5cm]{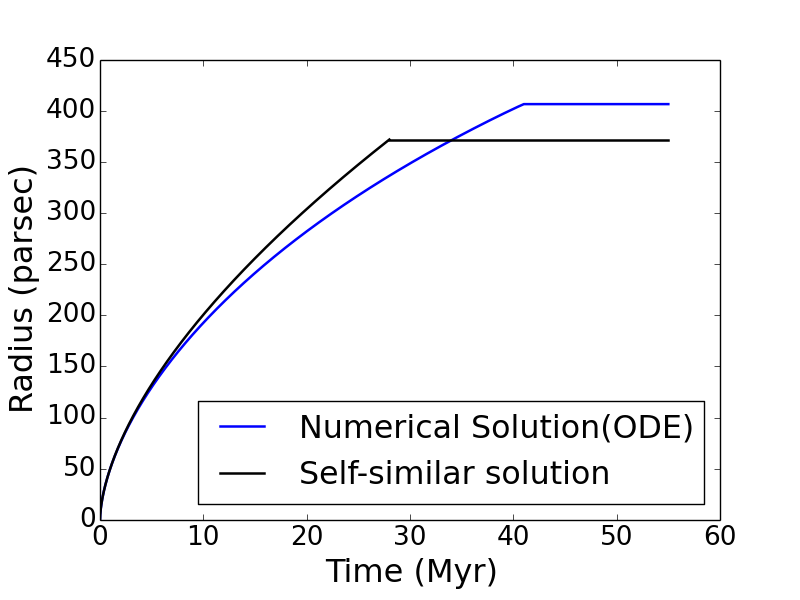} }}%
    \caption{Evolution of bubbles for ambient pressure $p_0 = 3\times10^{-12} dyne/cm^2$ and $p_0 = 5\times10^{-13} dyne/cm^2$,
    for average luminosity $L=1.2\times 10^{37}$ erg s$^{-1}$.}%
    \label{numerical1}
\end{figure}
\begin{figure}
   \centering
    \subfloat{{\includegraphics[width=4.5cm]{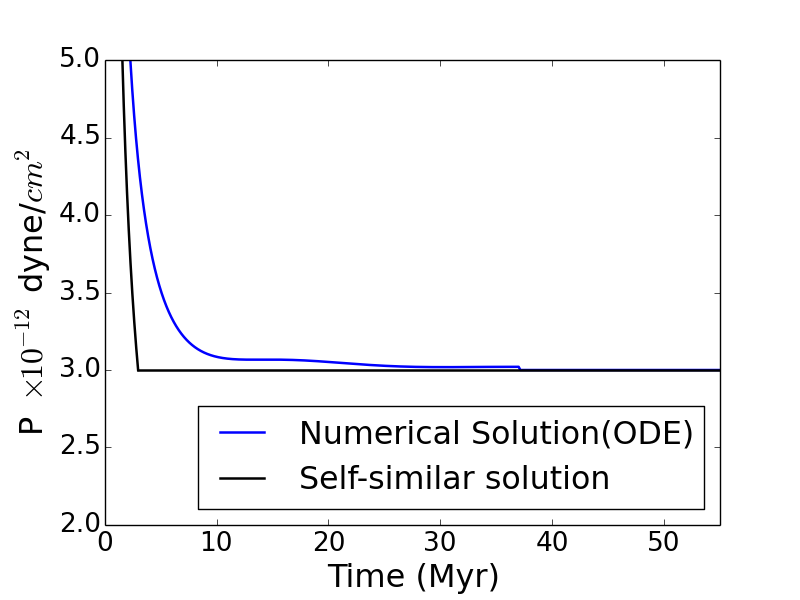} }}%
    \subfloat{{\includegraphics[width=4.5cm]{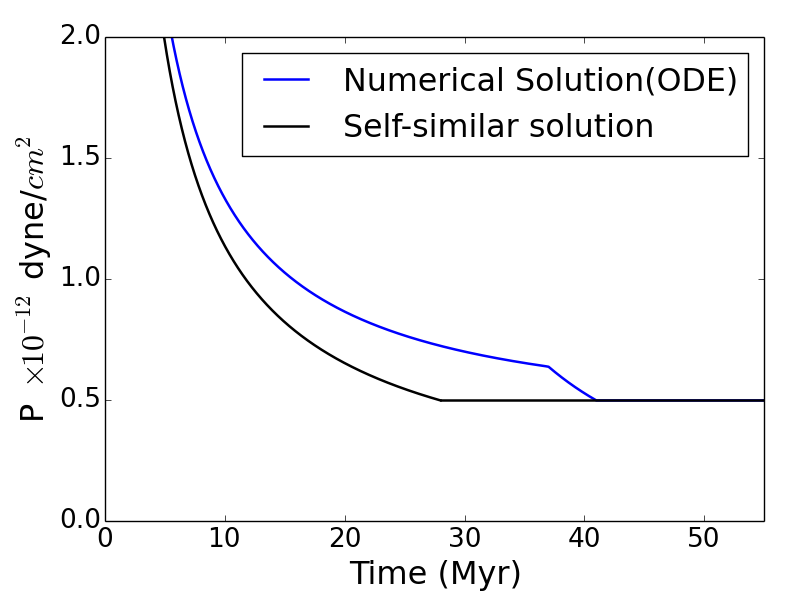} }}%
    \caption{Evolution of pressure inside bubbles for the same cases as in Figure \ref{numerical1}}%
    \label{numerical2}
\end{figure}

The wind bubble has four regions: a free wind region in which energy and mass are injected, a shocked wind region, a contact discontinuity and a shocked ISM region. The radius of the contact discontinuity evolves in the presence of an ambient pressure $p_0$, as,
(equations 54 and 56 in \cite{weaver1977})
\begin{eqnarray}
\frac{d}{dt}\bigg(\frac{4{\pi}}{3}R^3{\rho_0}\bigg) &=& 4{\pi}R^2(p-p_0)\,, \nonumber\\
\frac{dp}{dt} &=& {\frac{2}{3}}{\frac{L}{(4{\pi}/3)R^3}}-p\bigg(\frac{5  }{R}{dR \over dt}\bigg)
\label{eq-pressure}
\end{eqnarray}
Note that OC97 did not include $p_0$ in this relation, and only used $p_0$ to as
a criterion for stalling bubble growth.
We continue with the assumption of OC97 that the bubbles stall when the interior pressure becomes equal to the ambient pressure. The results
of eqn \ref{eq-pressure} show that the interior pressure decreases less rapidly than that assumed in OC97, and consequently, the bubble size grows to larger values than are admitted in self-similar evolution. We show two examples in Figure \ref{numerical1}, where the evolution of the bubble sizes in the self-similar case is compared with that determined from equations \ref{eq-pressure}, for two values of ambient pressure. The curves are for an average luminosity (given the above mentioned luminosity function), $L_{\rm av}=1.2 \times 10^{37}$ erg s$^{-1}$. We also show in Figure \ref{numerical2} the evolution of interior pressure in these two cases. The curves show that the interior pressure evolves slower than in the case of self-similar case, as mentioned above. 

\begin{figure}
   \centering
    \includegraphics[width=8cm, angle=0]{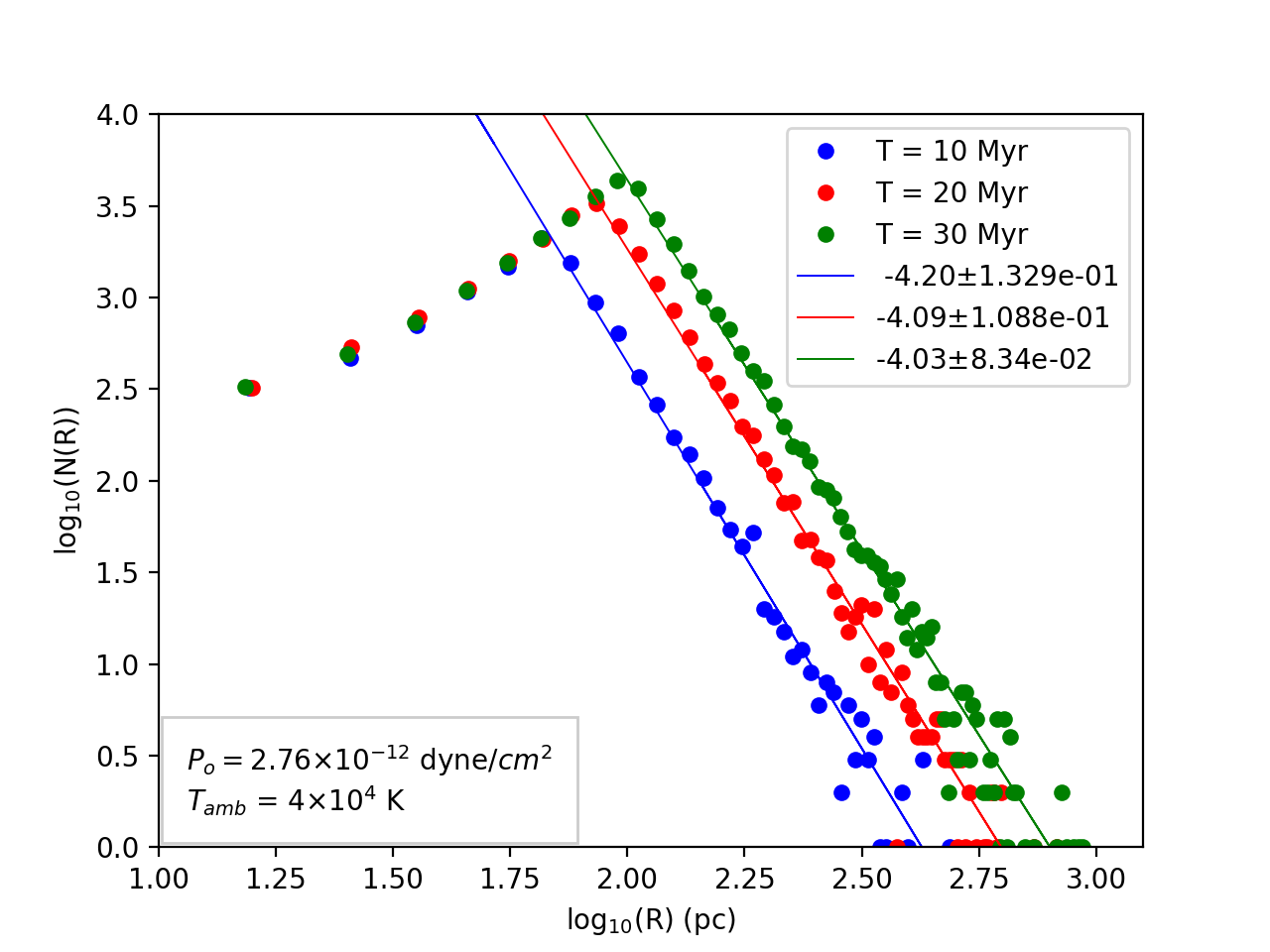}
    \caption{The size distribution of bubbles for adiabatic  evolution for an ambient pressure $p_0=2.76 \times 10^{-12}$ dyne cm$^{-2}$, at $10,\,20,\,30$ Myr. The distributions have a slope $\sim -4$, much steeper slope than 
    $-3$ in the case of self-similar evolution. Numerical values of the slopes are indicated in the diagram.}%
    \label{size-dist-ad}
\end{figure}

The corresponding distribution will be steeper than the self-similar case. 
This is shown in Figure \ref{size-dist-ad} for three different epochs. The slopes of the distribution is close to $\sim -4$, in this particular example, much steeper than $-3$ in the case of self-similar evolution.
The largest bubbles are mostly produced by massive clusters, with large mechanical luminosity, and for which the {revised relation for the ambient pressure (eqn \ref{eq-pressure}) does not make substantial difference to the bubble size evolution. The number of large bubbles therefore remain more or less the same, whereas the number of smaller bubbles grow to larger sizes than before. This will make the size distribution steeper than the self-similar case. 

The main reason for the steepness is the fact that growing bubbles dominate the population of superbubbles at any time, since the bubbles tend to grow for a longer time in this case. In the case of domination by growing bubbles, supposing a power-law scaling relation between radius and luminosity as $R\propto L ^x$, the slope of the size distribution is given by 
\be
N(R) \propto \phi(L) \Bigl ( {\partial R \over \partial L} \Bigr )^{-1} \propto R^{{1-x-\beta\over x}} \,.
\ee
For example, for self-similar solution of \cite{weaver1977}, $x=1/5$ and $N(R) \propto R^{4-5\beta}$, as shown by OC97. In our case, since the power-law index is close to $-4.1$, it implies $x \approx 1/3.1$, although there is no simple analytical scaling relation for radius and luminosity  for this case. We have confirmed this from our numerical results. 

However, the assumption of bubble stalling when pressure equilibrium is reached is not valid, since the bubble shell continues to move outward 
because of its momentum. This will continue to make the bubbles bigger in size until the outer shock speed becomes comparable to the sound speed of the ambient medium. This can be demonstrated with the help of numerical hydrodynamical simulations, which we describe below.

\section{Simulation results}
Here we include the effect of radiative cooling in the evolution of superbubbles. We have used \textit{PLUTO} for 1-D numerical hydrodynamical calculations \citep{mignone2007}. We have solved the following equations:
\begin{eqnarray}
\frac{{\partial}\rho}{\partial t} + \triangledown{\cdot}(\rho\textbf{v}) &=& S_m\nonumber\\
\frac{\partial(\rho\textbf{v})}{\partial t} + \triangledown{\cdot}(\rho\textbf{v}\otimes\textbf{v}) &=& -\triangledown p\nonumber\\
  \frac{\partial{e}}{\partial{t}} + {\triangledown}{\cdot}[(e+p)\textbf{v}] &=& S_e - q^-
 \label{hydro}
\end{eqnarray}
Here, total energy $e = \rho(\epsilon + \frac{1}{2}v^2)$, $\epsilon $  is specific energy, $\rho $ is mass density, $p$ is  pressure and \textbf{v} is the fluid velocity. The terms $S_m$ and $S_e$ correspond to the mass-loss rate $\dot{M}/V$ and input mechanical
energy ($L/V$) respectively, which are related to each other by the wind speed, $v_w$, as, $L=0.5 \dot{M} v_w^2$, and we assume $v_w=2000$ km s$^{-1}$ \citep{chevalier1985}.} We have introduced the source terms as per the model of \cite{chevalier1985}. We have kept the ambient particle density $(n_{\rm amb})$
and ambient temperature$(T_{\rm amb})$ constant for each run. Equations \ref{hydro} have been solved in 1D spherical coordinate using the HLLC solver (CFL Number = 0.3)

\begin{figure}
   \centering
    \subfloat{{\includegraphics[width=4.5cm]{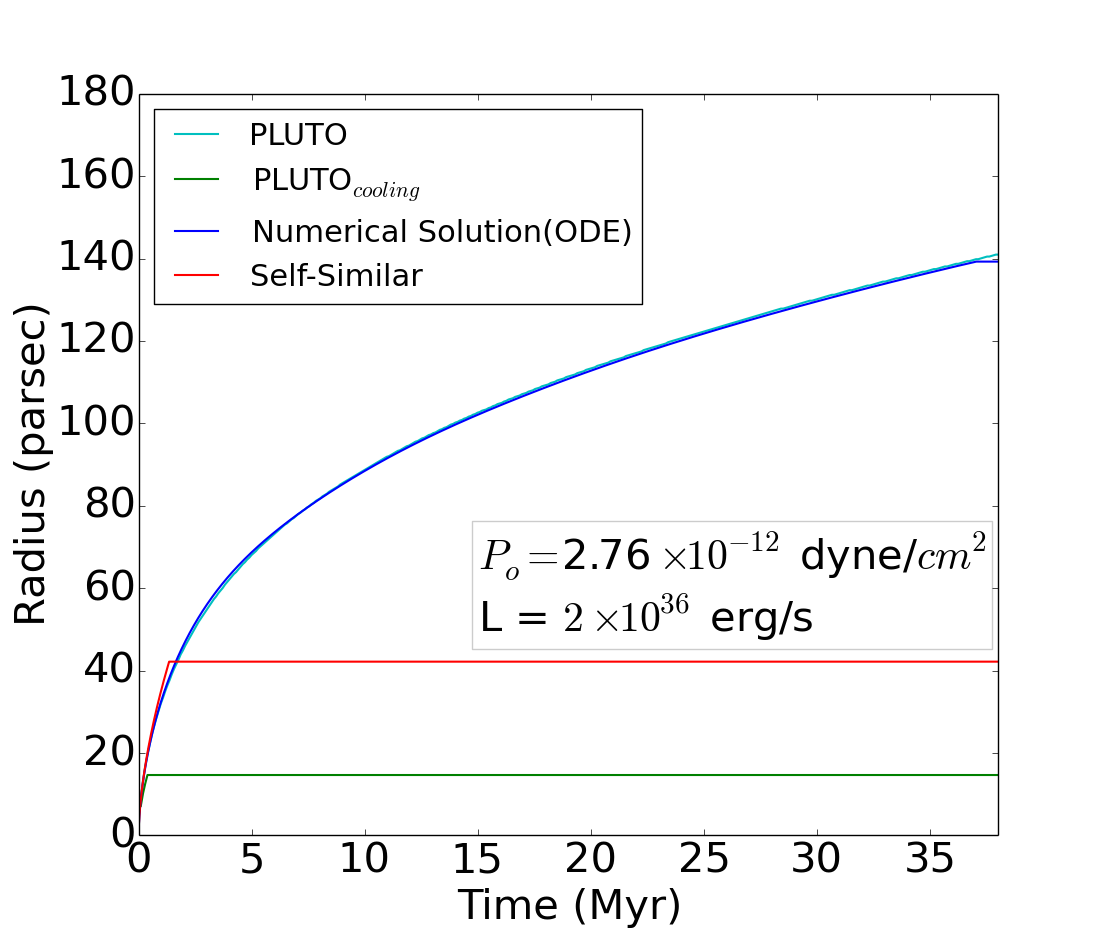} }}%
    \subfloat{{\includegraphics[width=4.5cm,height = 3.6cm]{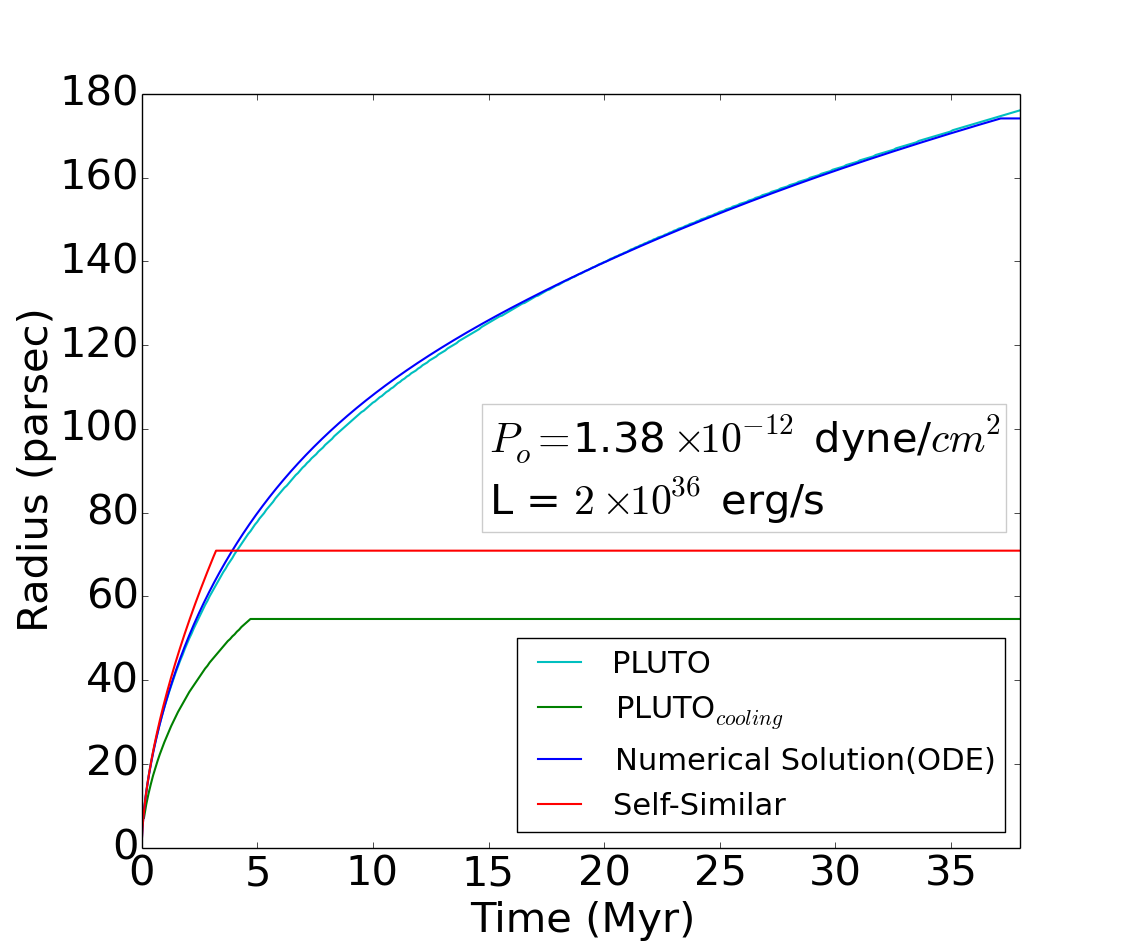} }}%
   \label{pluto1}
    \subfloat{{\includegraphics[width=4.5cm,height=3.6cm]{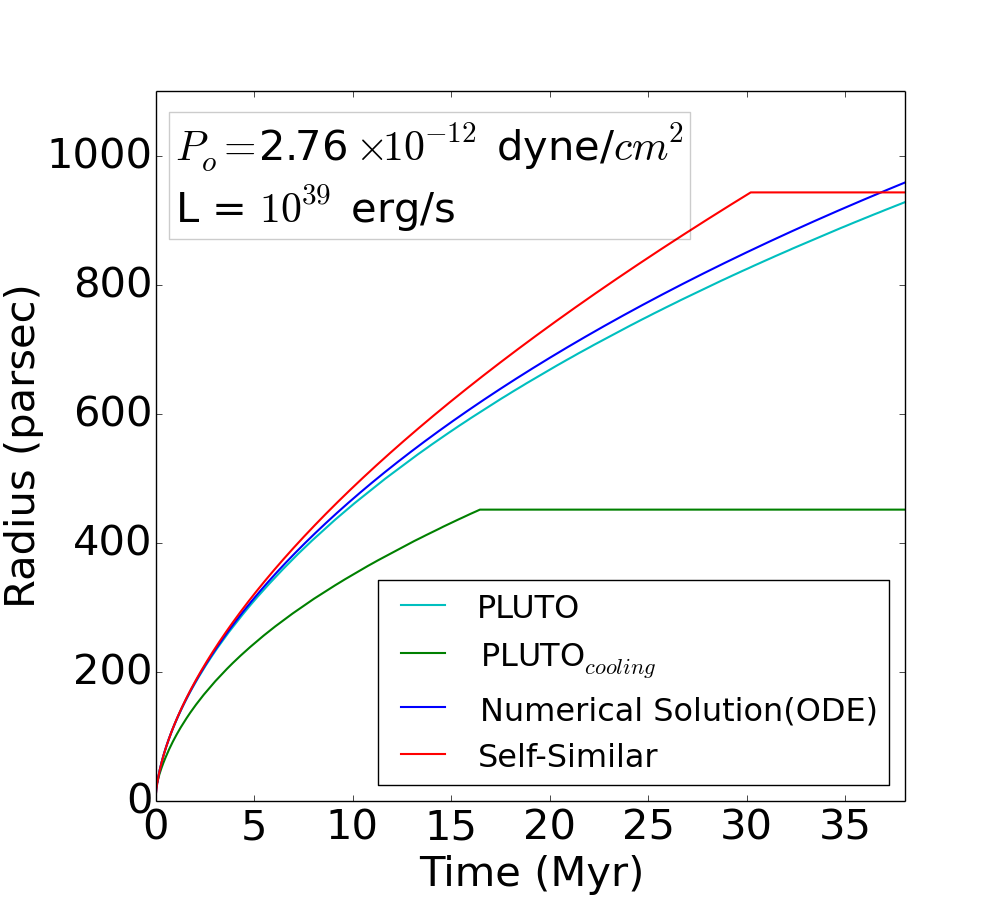} }}%
    \subfloat{{\includegraphics[width=4.5cm]{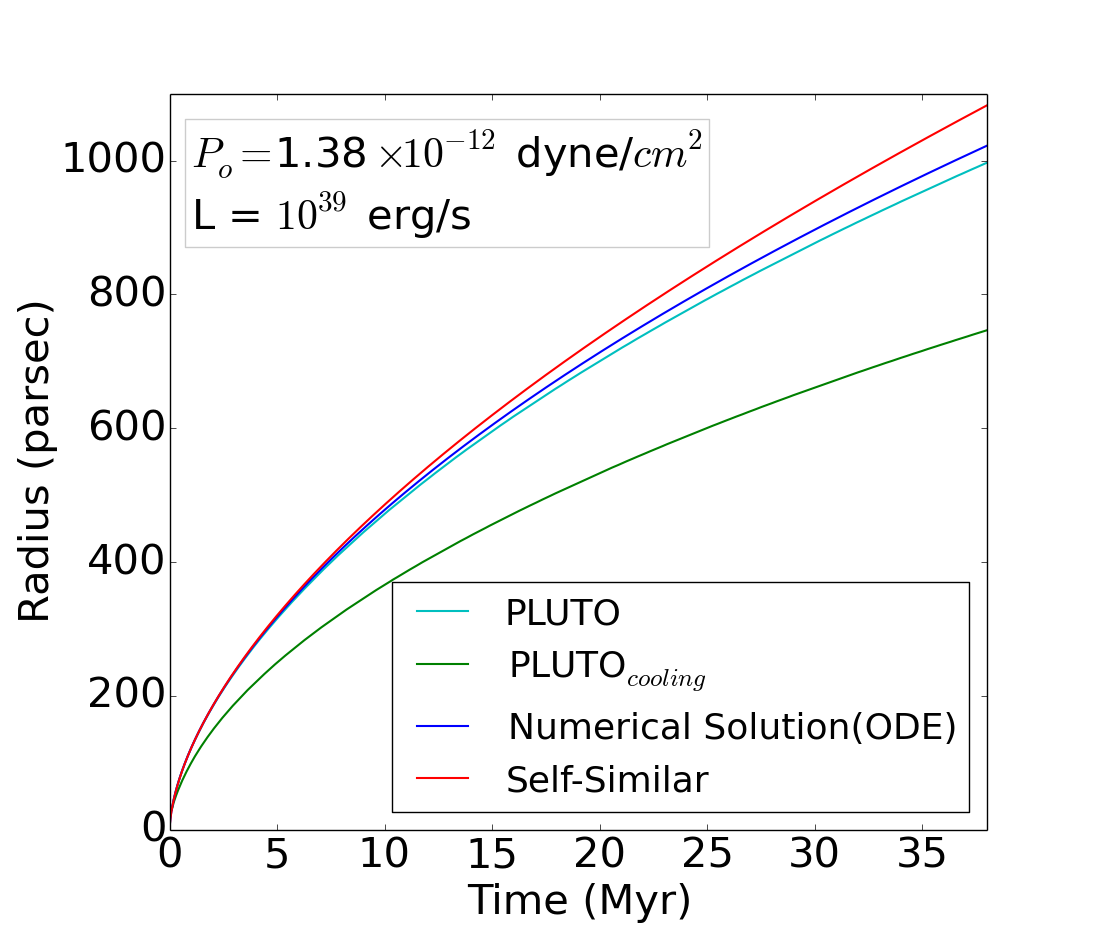} }}%
    \caption{Evolution of radius bubbles from hydrodynamical simulations is compared to the self-similar case. The upper two panels are for
    $L=2 \times 10^{36}$ erg s$^{-1}$ and lower panels, for $L=10^{39}$ erg s$^{-1}$. The left panels show the case for ambient pressure
    $p_0=2.76  \times 10^{-12}$ dynes cm$^{-2}$, and the right panels, for $p_0=1.36 \times 10^{-12}$ dynes cm$^{-2}$. Red curves plot the adiabatic self-similar evolution (assuming stalling at pressure equilibrium). Blue curves show the results of using eqn \ref{eq-pressure} (assuming stalling at pressure equilibrium). Cyan curves show the results of numerical simulation without cooling, and green curves show the results of
    simulation with cooling, assuming stalling when outer shock speed equals ambient sound speed.}%
 \label{pluto2}
\end{figure}

Since the appropriate medium for embedding the superbubbles is the warm neutral medium of ISM, we use $n_{\rm amb}=0.5$ cm$^{-3}$
for all our runs \citep{wolfire2003}. However, we consider the case of different equivalent ISM pressures by assuming different $T_{\rm amb}$. It is known that non-thermal pressures in ISM, especially in neutral medium that we are concerned with here,  can be substantial \citep{elmegreen2004} and even more than the thermal pressure. For example, in the solar neighbourhood, \cite{jenkins2011} estimated that turbulent and magnetic pressures are comparable in magnitude, and are roughly three times larger than the thermal pressure. In order to take into account the effect of non-thermal pressure, we use a corresponding equivalent ambient temperature, keeping $n_{\rm amb}$ fixed. We use  $T_{\rm amb}= 4\times 10^4$ K, for  corresponding ISM pressures $p_0= 2.8 \times 10^{-12}$ dynes cm$^{-2}$ for our fiducial runs, but have also varied the value of $T_{\rm amb}$ in order to study its influence on size distribution.


Mass and energy are continuously injected within a small radius $r_c$. We use $r_c=1$ pc, in order to minimize non-physical cooling losses at the early stages of shock formation (see eqn 4 in \cite{sharma2014}). According to this criterion, for the lowest luminosity considered here ($L=10^{36}$ erg s$^{-1}$), $r_c$ should be less than $\le 2.5$ pc.
Hence the source 
terms $S_m = \dot{M}/V_c$ and $S_e = L/{V_c}$ are non-zero for $r<r_c$, where $V_c = \frac{4\pi}{3}r_c^3$.
In the last equation for energy conservation, $q^{-} = n_i n_e \Lambda(T)$,  $\Lambda(T)$ being the cooling function. We have used a tabulated cooling function for solar  metallicity. We turn off cooling when temperature comes down to $10^4$ K, to mimic the effect of photoionization heating in the bubble. 

We show the results of simulations with and without cooling in Figure \ref{pluto2}. The luminosities used here is a low value of $L=2 \times 10^{36}$ erg s$^{-1}$ (top panels), and a high luminosity of $L=10^{39}$ erg s$^{-1}$ (bottom panels) . The left panels are for $T_{\rm amb}=4 \times 10^4$ K, and the right panels,  for $T_{\rm amb}=2 \times 10^4$ K. The self-similar case is show in red, and the result of using eqn \ref{eq-pressure} is shown for comparison in dark blue, although without the condition of stalling at equal pressure. The results of simulations with and without cooling are shown in green and cyan, respectively.






We first notice that in the case of no radiative cooling (cyan), the evolution of the shell is roughly similar to that of eqn \ref{eq-pressure}, except for high
mechanical luminosity, in which case the PLUTO runs give a slightly smaller radius (Figure \ref{pluto2}). This is because of the fact that eqns
\ref{eq-pressure} do not take into account mass injection, which is higher in our calculations for higher luminosities (since the wind speed is considered equal in all cases). The injected mass increases the inertia of the shell  and decelerates it to some extent.


Secondly,  increasing the ambient pressure (without increasing the gas density) makes the bubbles smaller, as expected physically.

\subsection{Adiabatic case}
\begin{figure}
\centering
\includegraphics[width=8cm, angle=0]{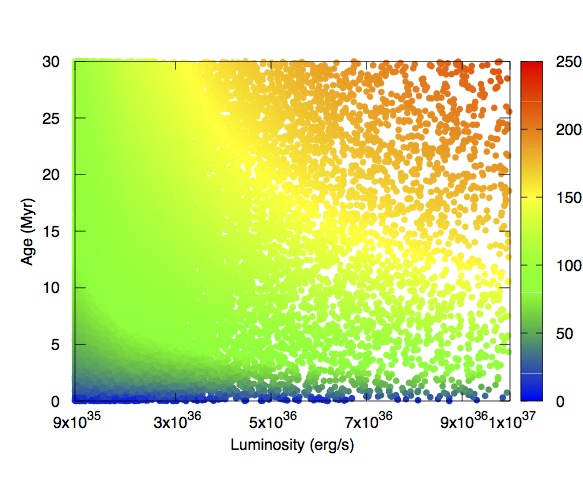}
\caption{Scatter plot of bubbles in the parameter space of mechanical luminosity $L$ and bubble age, without cooling, after 30 Myr of the onset of 
star formation, in which the colour of the
data points refer to bubble sizes, as shown in the colour palette on the right.
}
\label{scatterplot1}
\end{figure}

Let us first discuss the size distribution in the adiabatic case. As mentioned above, the results of the simulations confirm our analytical results in \S 3, where the pressure gradient term is included in the dynamics of bubbles. We had seen in that case that size distribution is steeper than $-3$, because of domination of growing bubbles. This is confirmed by the simulation results in which cooling is turned off, and we get a size distribution for the fiducial case with a power-law index $\approx -4$. The scatter plot of bubbles in this case is shown in Figure \ref{scatterplot1} as a function of mechanical luminosity (up to $10^{37}$ erg s$^{-1}$) and bubble age, for a snapshot at 30 Myr. The size of the bubbles are shown in different colours according to the colour palette shown on the right, the red ones being the largest and blue ones being the smallest. It can be seen that the envelopes for different colours (which can be thought of iso-size contours) delineate curves lines in the parameter space. Consider the self-similar case for a moment, in which the combination $Lt^3$ appear together in the relation for size $R \propto (Lt^3)^{1/5}$. If bubbles were to grow with this scaling, then these envelopes would be curves of $t \propto L^{-1/3}$. But these curves in Figure \ref{scatterplot1} are more complicated, signifying non-self-similar evolution of bubbles, even in the adiabatic case, because of the presence of pressure gradient term that is neglected in the \cite{weaver1977} solution.

\subsection{Effects of cooling}
Let us now discuss the effects of cooling.  We notice from Figure \ref{pluto2} that the inclusion of radiative cooling leads to a large difference in the evolution of the shell radius. 

\begin{figure}
   \centering
    \includegraphics[width=9.2cm, angle=0]{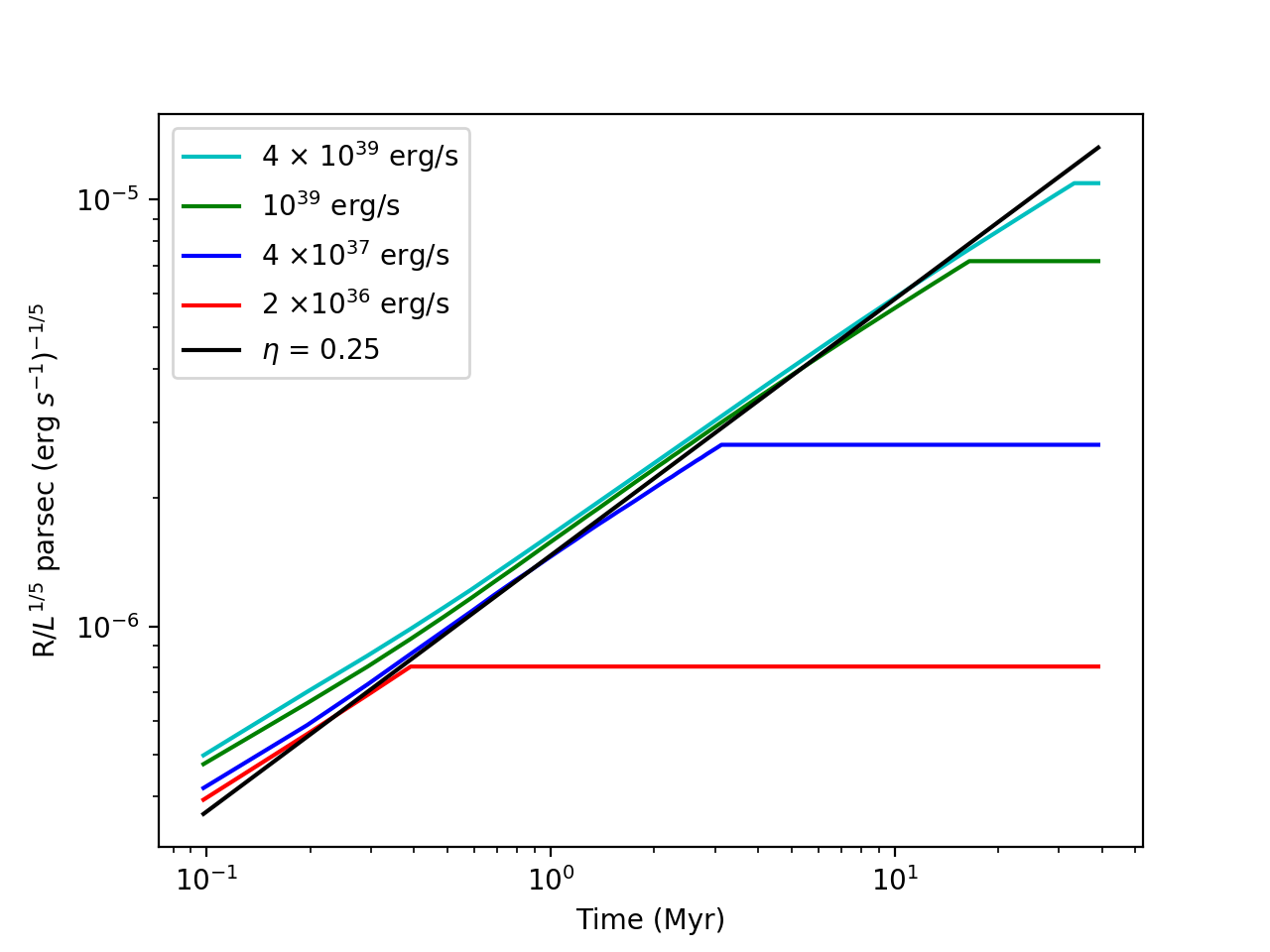}
    \caption{Logarithm of the ratio of radius to mechanical luminosity $L$ in bubbles (in units of cm/(erg s$^{-1}$)) is plotted against time (in Myr), for four different values of $L$, shown in different colours. The ambient pressure is $P_{\rm amb}=2.76 \times 10^{-12}$ dyne cm$^{-2}$, and equivalent temperature is $4 \times 10^4$ K. The near superposition of the cases of all luminosities show that a single value of $\eta$ is able to describe the evolution of bubbles in the presence of radiative cooling. The factor $\eta$ is found to be $\approx 0.25$, making the bubbles a fraction $\approx 0.76$ times smaller than their adiabatic case. }%
    \label{eta}
\end{figure}

The evolution of the bubbles of different luminosities in the presence of cooling and the pressure difference between ambient and gas inside the bubble, can be roughly described by a single parameter $\eta$, which can take into account radiation loss, where $\eta$ is defined in terms of $R=0.76 (\eta L t^3/\rho)^{1/5}$ (the pre-factor $\big({250 \over 308 \pi}\big)^{\frac{1}{5}}\approx 0.76$ applies to the adiabatic case \cite{weaver1977}). We show in Figure \ref{eta} the ratio of radius to $L^{1/5}$ versus time for bubbles of four different luminosities. The curves show that they are nearly superposed on one another, which indicates that a single value of $\eta$ can describe their evolution, whose value in this case (for the choices of ambient conditions) is inferred to be $\sim 0.25$. Similar conclusions have been reached by previous workers, e.g., \cite{maclow1988, krause2014}. More recently, \cite{sharma2014} showed that bubbles can retain a fraction of the input energy, of order $0.2\hbox{--}0.4$ depending on ambient conditions (their Figures 7 and 8). They also showed that this conclusion is valid even in the presence of thermal conduction. \cite{yadav2017} further showed that this fraction decreases with increasing ambient density ($\eta \propto n_{\rm amb}^{-2/3}$) and is of order $\sim 20\%$ for ambient density of $0.5$ cm$^{-3}$ (their Figure 8), consistent with our estimate in the present work. Figure \ref{eta} shows that at early epochs, the value of $\eta$ can vary with luminosity, since the curves for different luminosities do not quite overlap. However, they roughly do so by the time of bubble stalling, which is more relevant to our present work. In general, bubbles with lower luminosity suffer more loss of energy due to radiation (or, $\eta$ is smaller for lower $L$), although this trend is reversed below $4 \times 10^{37}$ erg. 
This is because of the fact that 
the outer-shock speed is lower for low luminosity shells, and the resulting post-shock temperature is also reduced, leading to a higher rate of radiative loss, since the cooling function in the relevant temperature range is a decreasing function of temperature. At the lowest luminosities considered here, the post-shock temperature is close to $10^4$ K even at the earliest epochs, where the cooling function drops, reversing the trend.

We also notice that, as argued previously, bubbles do not stall when pressure inside becomes equal to the ambient pressure. However, the
continuation of expansion seen in the simulation is also misleading, because our use of equivalent temperature for non-thermal pressure of the ambient medium has limitations. The non-thermal pressure in the ambient medium due to turbulence and other process will fragment or distort the shell, robbing it of its momentum which would have otherwise make it expand further. This is not easy to simulate without initially introducing density inhomogeneities and turbulence in the ambient medium in the numerical set-up, which would increase the number of free parameters in the calculations. The fact that shells would not be able to retain their identity when their nominal speed becomes comparable to the ambient medium sound speed is self-evident.  For example, for supernovae blast waves, this was the condition imposed by \cite{mckee1977} in their three phase model of ISM.

Therefore, 
we impose a condition of stalling the shell when the outer shock speed equals the (isothermal) sound speed of the ambient medium. 
This is shown by the horizontal section of the green curves in Figures \ref{pluto2}.



\begin{figure}
\centering
\includegraphics[width=8cm, angle=0]{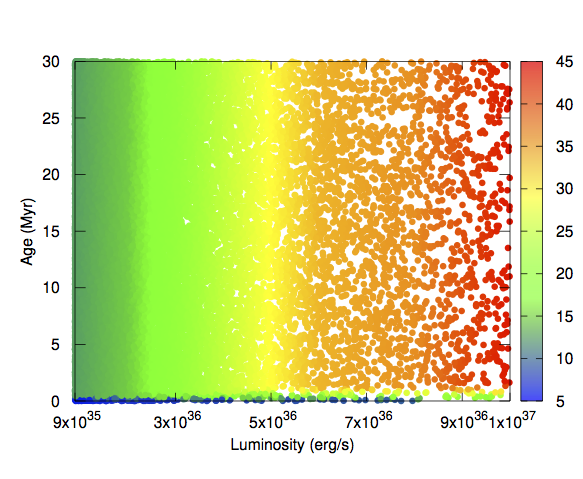}
\caption{Scatter plot of bubbles in the parameter space of mechanical luminosity $L$ and bubble age, after 30 Myr of the onset of 
star formation, in which the colour of the
data points refer to bubble sizes, as shown in the colour palette on the right.
}
\label{scatterplot}
\end{figure}


\section{Results}
We first show in Figure \ref{scatterplot} a scatter plot of the bubbles as a function of mechanical luminosity (up to $10^{37}$ erg s$^{-1}$) and bubble age, for a snapshot at 30 Myr after the
onset of star formation. The bubble sizes being shown in different colours according the colour palette shown on the right, the red ones being the largest and blue ones being the smallest. The star formation rate
is considered to be uniform and equal to $1$ M$_\odot$ yr$^{-1}$, and the luminosity function of clusters is assumed to obey a power-law with
index $\beta=2$. If we take a snapshot at 30 Myr for bubbles triggered by mechanical luminosity up to $10^{37}$ erg s$^{-1}$, then these are the bubbles that would show up. They will have different ages, as shown by their distribution along  the vertical axis. the vertical colour stripes indicate that most bubbles have stalled, being at the same radius at different times, except for the bubbles in the bottom of the figure.
This is in contrast to the case without cooling, in which the population of superbubbles is dominated by growing bubbles, in Figure \ref{scatterplot1}.

The scatter plot shows that at any given time (here, at 30 Myr),  the smallest bubbles are produced mostly by low luminosity clusters (blue circles) {\it and}
they are predominantly young. This is shown by the fact that blue dots mostly appear at the left bottom corner of this plot. The largest bubbles are created by clusters at the high luminosity end and can be both young and old. 
Most of the points in the scatter plot, however, arise due to stalled bubbles.
Leaving aside the smallest bubbles (blue), there are vertical columns
of different colours (different sizes) in the figure. This implies that the sizes of the bubbles mostly correspond to the mechanical luminosity of the bubbles, and almost independent of the age. This is because of the stalling condition we have imposed at the time of outer shock speed becoming comparable to the ambient medium sound speed.

\begin{figure}
\centering
\includegraphics[width=8cm, angle=0]{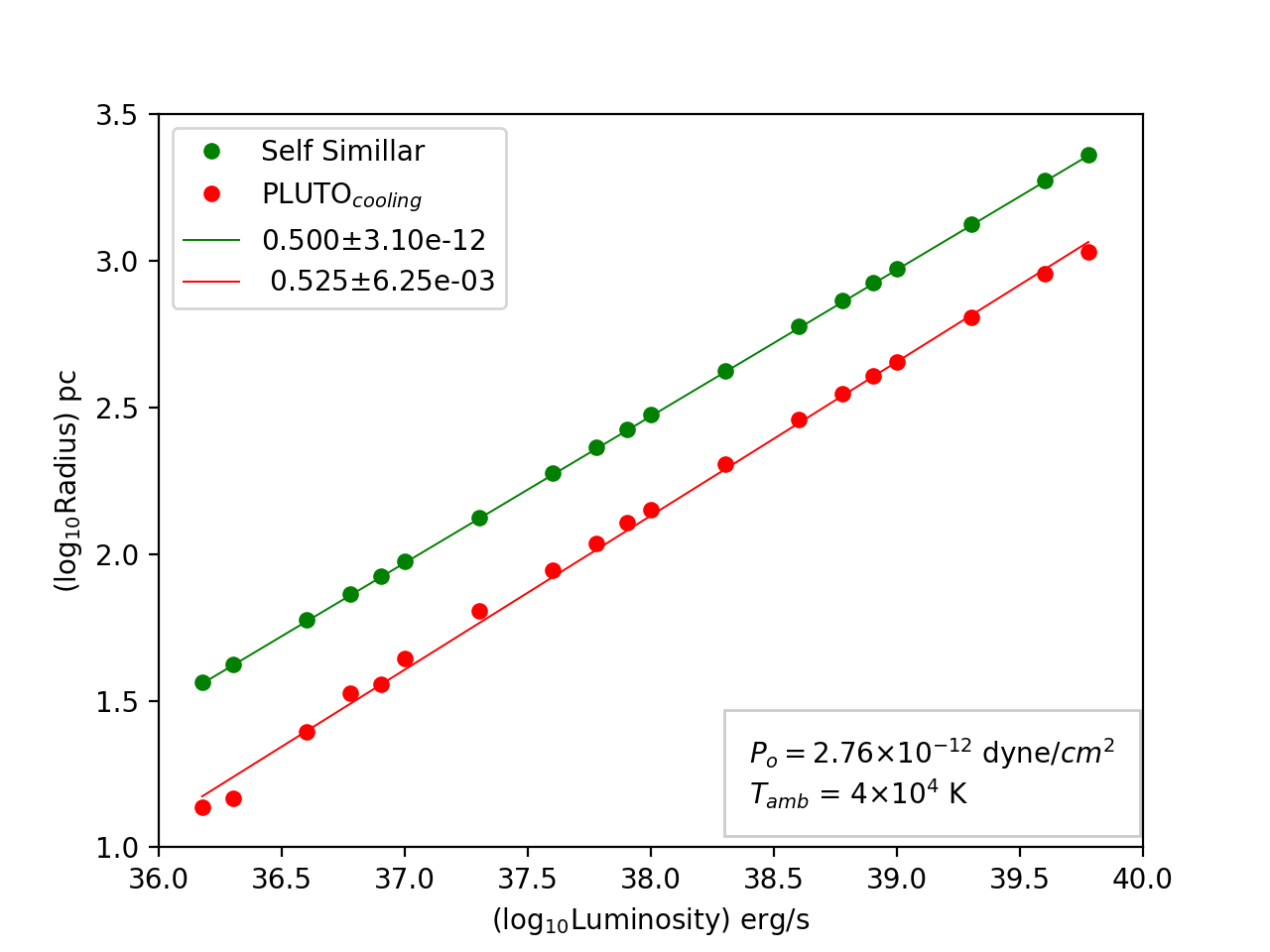}
\caption{Final bubble size as a function of mechanical luminosity, in the case of self-similar evolution (where stalling is assumed
to occur at pressure equilibrium stage, OC97), and from simulations with cooling (in which stalling is imposed when outer shock
speed equals the ambient isothermal sound speed).
}
\label{rvsl}
\end{figure}

The stalled radii can also be written in terms of the stalling time $t_s$ as  (equating the ambient isothermal sound speed of $\sqrt{P_0/\rho}$
with the outer shock speed, ${3 \over 5} R/t$),
\be
R_{cool,e}={5 \over 3} P_0 ^{1/2} \rho^{-1/2} \, t_s\,, 
\ee
and the corresponding luminosity is given by,
\be
L_e=\Bigl ({5 \over 3}\Bigr )^{5} \, \Bigl ({250 \over 308 \pi} \Bigr )^{-5} \,P_0 ^{5/2} \rho^{-3/2} \, t_s^2 \eta^{-1/2}\,.
\ee
These expressions allow us to estimate the largest bubble size $R_e$ that is reached after $t=t_s=t_e\approx 40$ Myr, after which the OB stars
in a cluster drop off the main sequence and the mechanical luminosity ceases to power any further growth of the bubble. For the same fiducial ISM parameters as above, we get $R_{cool,e}=1388.4$ pc (in the notations of OC97) and the corresponding luminosity is $L_e\approx 6 \times 10^{39}$
 erg s$^{-1}$. Clusters with luminosity larger than this will continue to grow and not stall even at $t_e (40)$ Myr. We find that these values are
 similar to the ones considered in OC97 ($R_e=1300$ pc, $L_e=2.2 \times 10^{39}$ erg s$^{-1}$). These estimates also give us an idea of 
 bubbles that are still evolving and have not reached steady-state at a certain epoch. For example, for the same ISM parameters, at $10$ Myr (after the onset of star formation), bubbles smaller than $347$ pc (corresponding to $L\sim 3.8 \times 10^{38}$ erg s$^{-1}$) have reached steady-state and larger ones are still  evolving.
 
 \begin{figure}
\centering
\includegraphics[width=8cm, angle=0]{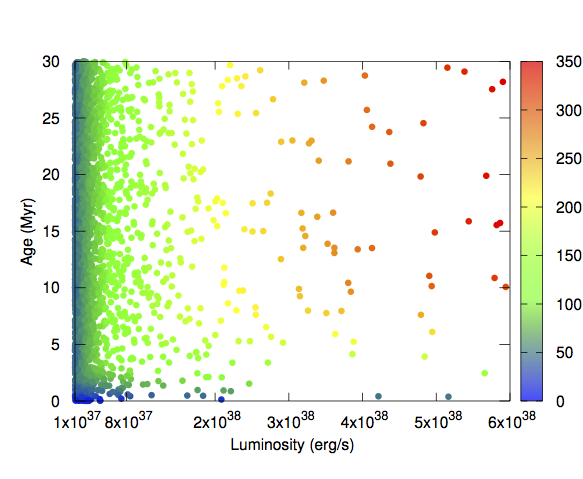}
\caption{Same as in Figure \ref{scatterplot1}, but for luminosities in the range  $10^{37}\hbox{--}6 \times 10^{38}$ erg s$^{-1}$.
}
\label{scatterplot-full}
\end{figure}
 
Figure \ref{rvsl} shows the relation between final bubble size from our simulations (red points) as a function of mechanical luminosity
and shows that the size scales as $L^{1/2}$. This follows from the fact that radius of a bubble scales as $R \propto (\eta L)^{1/5} t^{3/5}$,
where $\eta \sim 0.25$ takes into account the energy loss by radiation and pressure gradient, as mentioned earlier. 
This implies an outer shock speed
$v \propto (\eta L)^{1/5} t^{-2/5}$. Since the bubbles are assumed to stall when this speed becomes comparable to the ambient sound
speed, the stalling time scales as $t_s \propto L^{1/2}$, and consequently, $R_f\propto L^{1/2}$.

\begin{figure}
\centering
\includegraphics[width=9.6cm, angle=0]{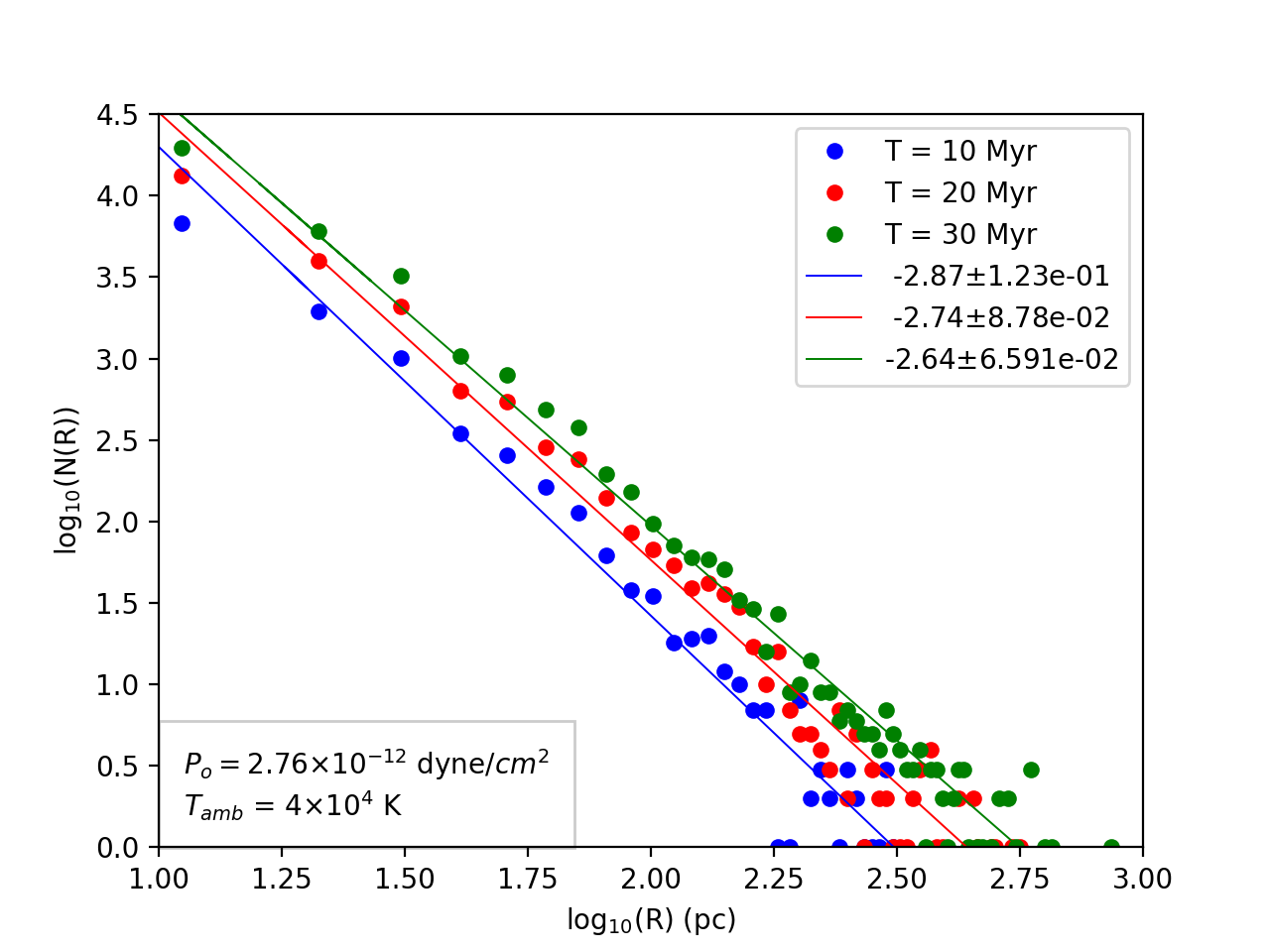}
\caption{Size distribution of bubbles in our simulation, for ambient pressure $p_0=2.76 \times 10^{-12}$ dynes cm$^{-2}$, at 10, 20, 30 Myr. The distributions at different epochs are fitted with power law, and the power-law indices are found to be $\sim 2.7$.
}
\label{size-simulation}
\end{figure}

This stalling condition is  similar to that used by OC97, since the ram pressure of the outer shock is related to the pressure of the
shocked wind region in a bubble \citep{weaver1977}, which dominates the pressure inside a bubble. Therefore, it is not a surprise that we get a similar relation between radius and luminosity as in OC97. However, the magnitudes of these two types of pressure are different. 
The inner pressure is roughly $\sim 0.8$ of the ram pressure \citep{Gupta2018}. Therefore when the inner pressure comes to equilibrium with ambient pressure, the
forward shock speed is still higher than the ambient sound speed. The stalling criterion based on speed therefore yields a larger radius by a factor
$\sim 1.2$. However when radiative cooling is taken into account, then this criterion gives a smaller radius because of radiation loss.

In the OC97 case, the stalled size is given by (using their equations 31 \& 32),
\ba
R_{f,OC}&\approx &{5 \times 7^{1/4}\over \sqrt{550 \pi}} \, P_0^{3/4} (\mu m_H \, n)^{1/4} \, L^{1/2} \nonumber\\
&\approx& 305 \, {\rm pc} \, \Bigl ( { P_0 \over 2.76 \times 10^{-12} \, {\rm dyne} \, {\rm cm}^{-2}} \Bigr )^{-3/4} \, \nonumber\\
&& \times
\Bigl ( {n \over 0.5 \, {\rm cm}^{-3}} \Bigr )^{1/4} \, \Bigl ( {L \over 10^{38} \, {\rm erg} \, {\rm s}^{-1}} \Bigr )^{1/2} \,.
\ea
This relation is shown by the green line in Figure \ref{rvsl}. For the case of cooling  in our simulation, equating the shock speed with the ambient 
(isothermal) sound speed leads to,
\ba
R_{cool}&\approx &\Bigl ({5 \over 3}\Bigr )^{-3/2} \, \Bigl ({250 \over 308 \pi} \Bigr )^{1/2} \, P_0^{-3/4} (\mu m_H \, n)^{1/4} \, (\eta L)^{1/2} \nonumber\\
&\approx& 180 \, {\rm pc} \, \Bigl ( { P_0 \over 2.76 \times 10^{-12} \, {\rm dyne} \, {\rm cm}^{-2}} \Bigr )^{-3/4} \, \nonumber\\
&& \times
\Bigl ( {n \over 0.5 \, {\rm cm}^{-3}} \Bigr )^{1/4} \, \Bigl ( {L \over 10^{38} \, {\rm erg} \, {\rm s}^{-1}} \Bigr )^{1/2} \, \Bigl ({\eta \over 0.25}\Bigr)^{1/2}\,.
\label{rcool}
\ea

The determination of the outer shock speed and the stalling time, however, involves the smoothening of the outer shock speed versus time, and 
the final determination of the stalled radius has an uncertainty of order $\lesssim 20\%$ owing to this. 
The stalled radii in the case of cooling are shown with red points in Figure  \ref{rvsl}. 

The second reason why we get a similar scaling between radius and luminosity is the fact that
radiation and ambient pressure affects the size evolution through a single factor $\eta$, leaving the dependency  of size on luminosity the intact. These two facts contrive to make the results in OC97 and in the present work look similar, even in the presence of cooling. 

\begin{figure*}
   \centering
    \subfloat{{\includegraphics[width=8cm]{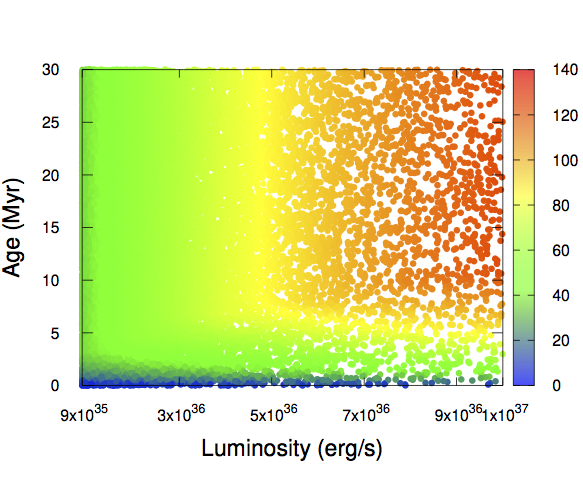} }}%
    \subfloat{{\includegraphics[width=8cm]{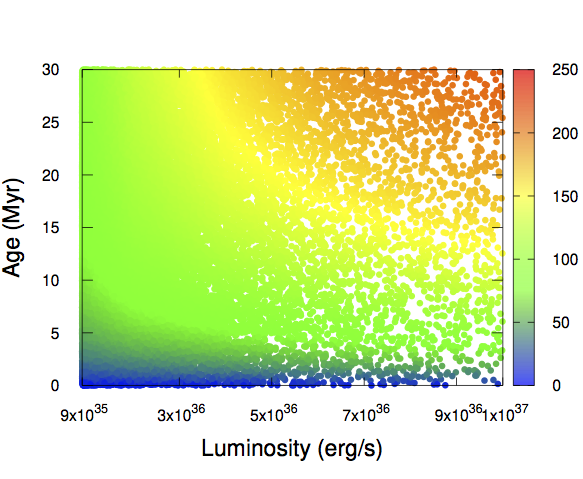} }}%
    \caption{Scatter plot for bubbles for ambient pressure $p_0=1.38 \times 10^{-12}$ dynes cm$^{-2}$ (left), and $6.9 \times 10^{-13}$ dynes cm$^{-2}$ (right), at  30 Myr. }%
    \label{scatterplot2}
\end{figure*}

The corresponding size distribution is, therefore, again expected to of the type $N(R) \propto R^{1-2\beta}$, and it is shown in Figure \ref{size-simulation}.
The fitted power-law indices at different epochs (10, 20, 30 Myr) are roughly $\sim 2.7\hbox{--}2.9$. This is  close to the value of $1-2\beta (=-3)$,
and the difference from $-3$ stems from the fact that $\eta$ weakly depends on luminosity. Had $\eta$ been totally independent of $L$, then, the size distribution would have been exactly $1-2\beta (=-3)$. Since $\eta$ is somewhat smaller for lower $L$ than its fiducial value, the shells for low luminosity clusters at stalled phase are somewhat smaller (see equation \ref{rcool}), leading to a somewhat flatter size distribution than the fiducial $1-2\beta$ value.
Since the radii of bubbles have decreased in general by a factor of $1.7$, and the peak radius has decreased by a similar  factor compared to the adiabatic case in Figure \ref{weaver}, to $\sim 10$ pc. While determining the size distribution we did not consider holes smaller than $10$ pc, and therefore we do not see any rising part in the distribution with radiative cooling.

One notes in Figure \ref{size-simulation} that the slope slowly changes with time, and becomes flatter. This is because of the fact that bubbles with large luminosity take a while to stall, and in any given snapshot, there would be newly formed (large luminosity) bubbles which would be in a growing phase. The stalling timescale is given by
\ba
t_s &\approx& 9.2 \, {\rm Myr} \Bigl ( { P_0 \over 2.76 \times 10^{-12} \, {\rm dyne} \, {\rm cm}^{-2}} \Bigr )^{-5/4} \, \nonumber\\
&& \times
\Bigl ( {n \over 0.5 \, {\rm cm}^{-3}} \Bigr )^{3/4} \, \Bigl ( {L \over 2 \times 10^{38} \, {\rm erg} \, {\rm s}^{-1}} \Bigr )^{1/2} \, \Bigl ({\eta \over 0.25}\Bigr)^{1/2}\,.
\label{tstall}
\ea
This implies that over time, the number of large luminosity bubbles, or consequently, large size bubbles would increase, whereas the small bubbles would have stalled quickly and their numbers would more or less freeze. This is shown in the scatter plot for the bubbles, in Figure \ref{scatterplot-full} for the range of luminosities  $10^{37}\hbox{--}6 \times 10^{38}$ erg s$^{-1}$. The large luminosity bubbles are seen to be growing in this plot, and the time scale for attaining stalled phase for $L\sim 2 \times 10^{38}$ erg s$^{-1}$ is $\approx 9$ Myr (shown by the fact that the onset of vertical bright yellow line corresponding to this luminosity in the Figure), as expected from equation \ref{tstall}. This will make the size distribution evolve over time, and make it flatter with time, as is seen in Figure \ref{size-simulation}. 

We show in Appendix A that our results are robust with respect to numerical resolution, although we note that our simulation does not address the issues of turbulence and mixing that may affect radiative losses at different resolutions.

The above results pose question, whether or not one always expects stalled bubbles (and consequently, a size distribution with slope $\approx -2.7$) in the presence of cooling, or if this depends on the ambient pressure. In order to answer the question, we have run our simulations for two different pressures.

Figure \ref{scatterplot2} shows the scatter plot of bubbles (similar to Figure \ref{scatterplot1}  and Figure \ref{scatterplot} for our fiducial
ISM pressure) for two different values of ISM pressures, at 30 Myr.  Essentially, we have decreased the non-thermal contribution in the ambient pressure, by reducing the equivalent temperature from $4 \times 10^4$ K to $2\times 10^4$ K (left panel) and then to $10^4$ K (right panel).
Comparing with Figures \ref{scatterplot1} (adiabatic case) and \ref{scatterplot}, we find that the bubbles in these two cases are {\it not} dominated by stalled bubbles, as was the case for Figure \ref{scatterplot}. For the left panel, we find that bubbles with $L\sim 5 \times 10^{36}$ erg s$^{-1}$ stall after a time scale of $\sim 7$ Myr, and bubbles bigger than that at latter times. When the pressure is further lowered, in the right panel, all bubbles keep growing until 30 Myr, and the circumstance is similar to Figure \ref{scatterplot1}. We recall that, in
these cases of domination by growing bubbles, the size distribution is likely to be steeper than $1-2\beta$, as we confirm below.

\begin{figure*}
   \centering
    \subfloat{{\includegraphics[width=8cm]{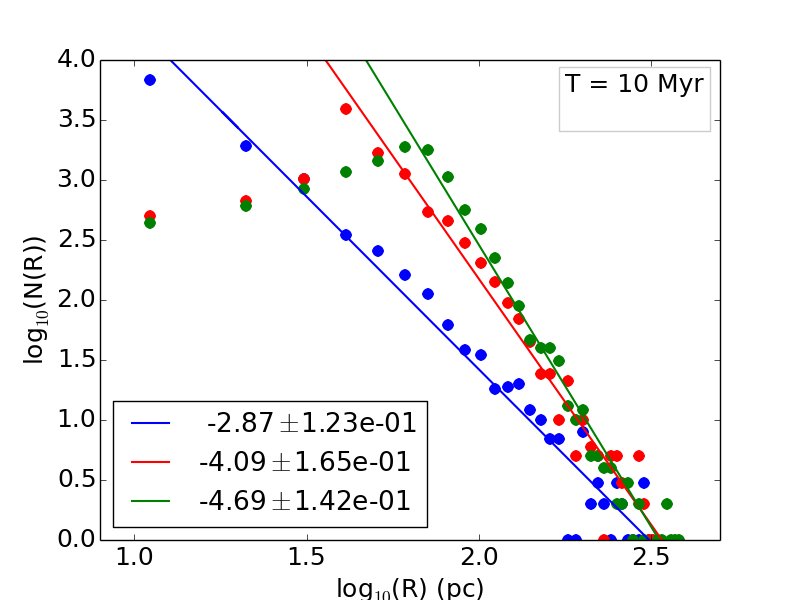} }}%
    \subfloat{{\includegraphics[width=8cm]{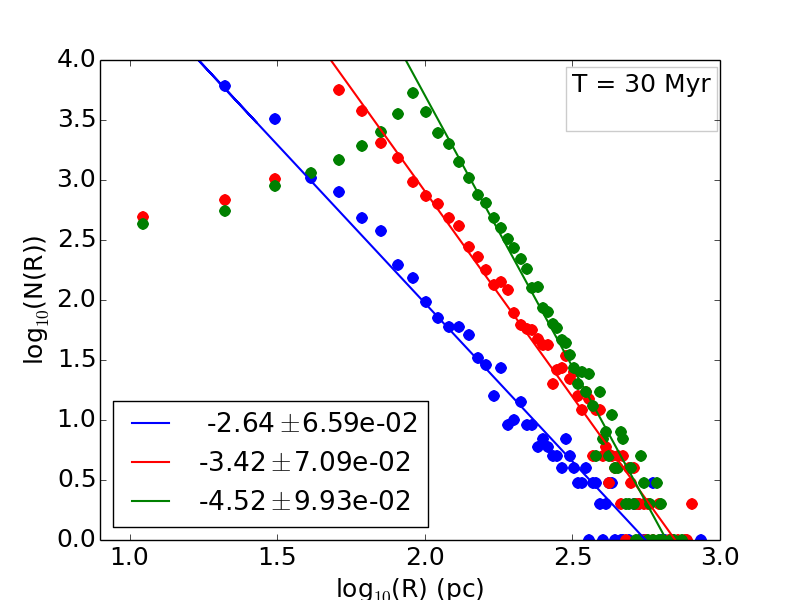} }}%
    \caption{Size distribution of bubbles for ambient pressure $p_0=2.76 \times 10^{-12}$ dynes cm$^{-2}$ (blue), $1.38 \times 10^{-12}$ dynes cm$^{-2}$ (red), $6.9 \times 10^{-13}$ dynes cm$^{-2}$ (green), at 10 and 30 Myr. }%
    \label{size-simulation2}
\end{figure*} 

\begin{figure}
\centering
\includegraphics[width=8.78cm, angle=0]{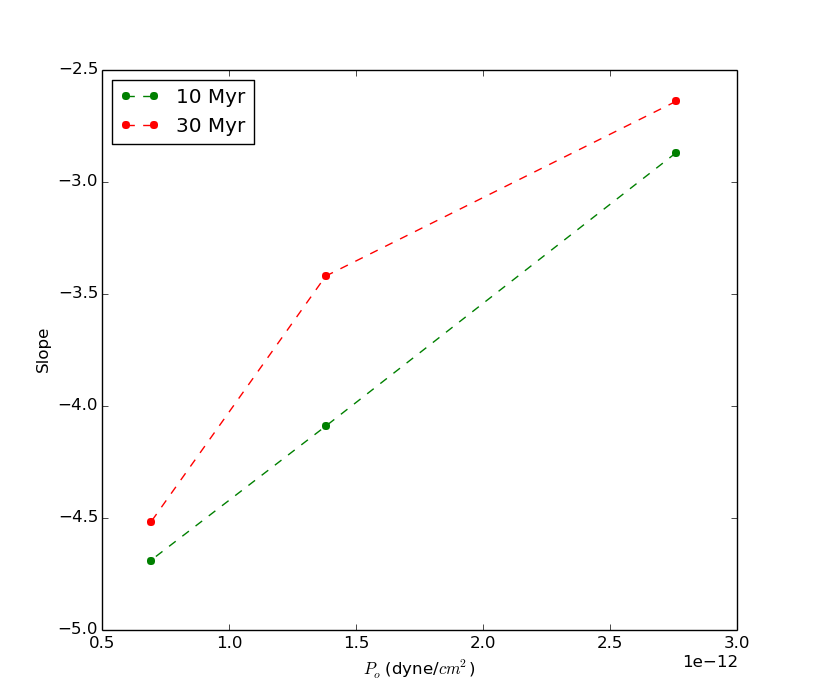}
\caption{The slope of size distribution as a function of ISM pressure, for two different epochs, 10 Myr (green) and 30 Myr (red). }
\label{slope-pressure}
\end{figure}

We show in Figure \ref{size-simulation2} the size distribution of the bubbles for these values of pressure at two different times after the onset of star formation (10 and 30 Myr). 
In general, we find that lowering pressure steepens the size distribution, by allowing low luminosity bubbles to grow to larger sizes. This trend is shown in Figure \ref{slope-pressure}, for two different epochs, 10 Myr and 30 Myr. At lower pressures, there is an evolution of the slope between these time scales, since bubbles keep growing in low ambient pressure, compared to high pressure environments.

The distributions at lower (non-thermal) pressures also show a positive
part at small sizes, in addition to the falling numbers at large sizes as we have seen earlier.
The rising part comes from 
growing young bubbles whose age is smaller than the stalling time of the bubble with the lowest luminosity. This  generates a peak in the distribution. The peak depends upon the radius of the stalled bubble with the lowest luminosity at the time being observed. Figure \ref{size-simulation2} shows that at higher pressures the peak radius is independent of time but starts varying with time  as soon as we start decreasing the pressure. This occurs due to the fact that at higher pressures (such as, at $2.76 \times 10^{-12}$ dynes cm$^{-2}$), the bubbles with  low luminosities stall much before 10 Myr hence the peak occurs at the same radius even when we observe the distribution at 30 Myr. However at lower pressures like $p_0=6.9 \times 10^{-13}$ dynes cm$^{-2}$ the bubbles with  low luminosities stall much after 30 Myr, and hence the peak shifts towards larger radii with time.  In other words, the deviation of the slope from $-3$ in low pressure environments indicate evolution of the bubbles, which breaks the relation of $L \propto R^2$ of stalled bubbles, which would have produced a $-3$ slope.

\section{Discussion}
It is known that observations support the $1-2\beta$ power law for the superbubble size distribution.  For example, \cite{bagetakos2011} found a power-law index of $\approx -2.9$ in {\it The HI Nearby Galaxy Survey}  of $20$ nearby galaxies.  This is consistent with $\beta \approx 1.9$.

They also reported a variation of the power-law index with galaxy type, with early type galaxies showing a steeper index ($\approx 4$) than late type galaxies. This indicates,
a preponderance of small size holes in early type galaxies. They associated this phenomenon with the level of star formation activity in different galaxy types. One obvious connection is that in early type galaxies, the observed mass function of star clusters is steeper than late type galaxies, as evidenced by the HII region luminosity function (e.g., \cite{kennicutt1989}). While the luminosity function in Sb-Sc galaxies has an index of $\sim -2$, the index varies from $\sim -1.7$ in Sc-Im galaxies \citep{banfi1993} to $\sim 2.6$ in Sa galaxies \citep{caldwell1991}.  \cite{oey1998} explained this variation in the HII region luminosity function in terms of a truncation in the maximum value of the luminosity distribution that depends on Hubble type.  This corresponds to early-type galaxies having a maximum cluster mass that is much lower than for late-type galaxies.

We mention in passing that it is also possible that lower star formation rate in early type galaxies would lead to lower non-thermal pressure, as has been seen in 
simulations \citep{joung2009}. According to their results, the turbulent pressure scales with surface density of star formation as $P_{\rm turb} \propto \Sigma_\ast^{2/3}$. Since our results indicate that decreasing non-thermal pressure steepens the size distribution, this remains another possibility. Future observations will be able to point towards the right explanation.

We note that our calculations do not take into account the merging of superbubbles and its effect on the size distribution. 
This was discussed by OC97 in terms of a porosity parameter $Q$ \citep{cox1974}, which is the ratio of superbubble volume to total ISM volume.  For the Milky Way, the SFR is near the critical point where $Q \sim 1$. 
It can be shown that the total volume occupied by superbubbles can exceed the volume of Milky Way ISM, considering a cylinder of 10 kpc radius and $500$ pc thickness, within $\le 1$ Myr, if the star formation rate is assumed to be $\sim 3$ M$_\odot$ yr $^{-1}$. Following \cite{clarke2002}, if one takes the average mass of clusters as $\sim 1300$ M$_\odot$, then the number of superbubbles produced per Myr is $\sim 2200$, for a star formation rate of $3$ M$_\odot$ yr$^{-1}$. Then with a size distribution with a slope of $-3$, the total volume exceeds the Milky Way ISM volume in $le 1$ Myr. This is also supported by the estimates of \citet{krause2015}. At the same time, the observed volume filling factor of HI shells in the Milky Way is less than $\sim 10\%$ \citep{bagetakos2011}. This implies that merging of superbubbles is important. It is also evident from the observations of \cite{simpson2012} that roughly $\sim 30\%$ of HI shells show signs of merging. OC97 also explore how the Milky Way $Q$ varies with the assumed value of $\beta$. Merging among superbubbles likely flattens the size distribution to some extent, by increasing the number of large bubbles at the expense of smaller bubbles. However, it is difficult to estimate the magnitude of this effect without a simulation that includes non-uniformity in the spatial distribution of star clusters.

\section{Conclusions}
We have studied the form and evolution of the size distribution of HI holes in the ISM of galaxies owing to superbubbles triggered by OB associations,
taking into account radiative cooling, with the help of numerical hydrodynamical simulations. 
 Previous works had assumed bubble growth stalls when the inner pressure of adiabatic bubbles equals  the ambient pressure, which is not valid since the bubbles maintain momentum-driven growth. 
Assuming that bubbles stall when the expansion speed becomes comparable to the ambient sound speed, we show that the inclusion of radiative cooling and ambient pressure
results in a power-law index of the size distribution  with slope $\sim -2.7$ for an ISM pressure of $p_0=2.8 \times 10^{-12}$ dyne cm$^{-2}$ and density $n=0.5$ cm$^{-3}$. This is consistent with observations by {\it THINGS}. 
We have further shown that decreasing the ISM pressure can make the population of growing bubbles dominating over stalled ones, consequently making the size distribution steeper. 
Our results imply that the size distribution can help interpret the evolution of bubbles, with the slope being $\sim -2.7$ in the case of
domination by stalled bubbles, and with a steeper slope for the case of growing bubbles. 
We have discussed the possibility that the size distribution in early type galaxies is steeper than late type galaxies because of the difference in the
intrinsic luminosity function as a function of galaxy type. A steeper luminosity function of star clusters in late type galaxies leads to a steep size distribution, as observed.

\bigskip
\section{Acknowledgement}
We would like to thank K. S. Dwarakanath, Nirupam Roy, Aditi Vijayan, Siddhartha Gupta and Ranita Jana for useful discussions. The authors thank the anonymous referee for helpful
comments. PD would like to thank Raman Research Institute for local hospitality during her visits for the project. 

\section{Appendix A: Convergence test}
\begin{figure}
\centering
\includegraphics[width=8.78cm, angle=0]{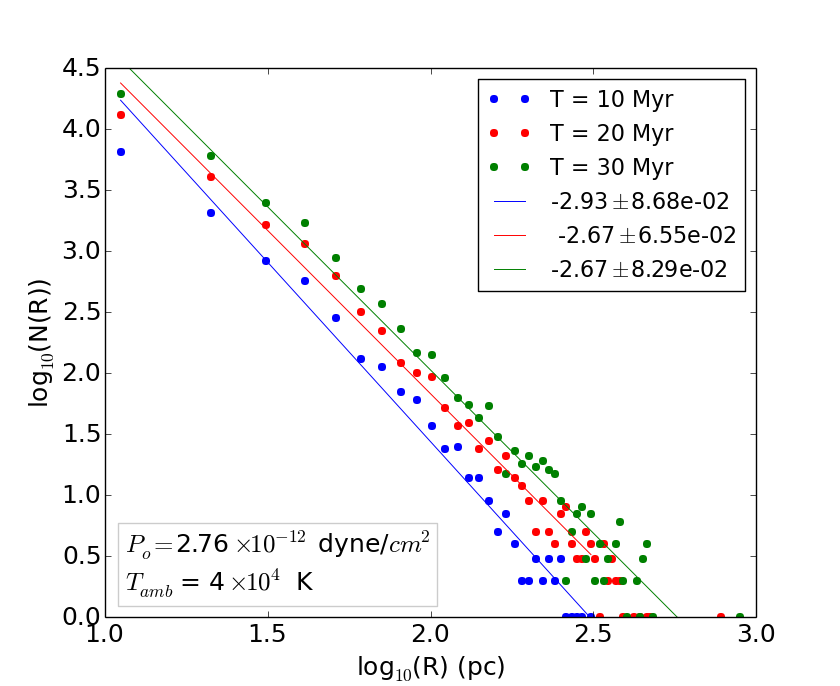}
\caption{Similar to Figure \ref{size-simulation} except that it is for $\Delta r=0.1$ pc.}
\label{convergence1}
\end{figure}

Here, we present the convergence tests of the numerical runs for superbubble evolution for our
fiducial case, for a different spatial resolution, with $\Delta r=0.1$ pc, instead of $0.16$ pc used
earlier. Figure \ref{convergence1} shows the size distribution  and fitted slopes at $10,20,30$ Myr, for the 
fiducial case of $P_0=2.76\times 10^{-12}$ dyne cm$^{-2}$. The slopes ($-2.93,-2.67, -2.67$) are similar
to the ones ($-2.87, -2.74, -2.64$) obtained for a coarser resolution (Figure \ref{size-simulation}). This confirms
the convergence of our results with respect to numerical resolution.

\begin{figure}
\centering
\includegraphics[width=8.78cm, angle=0]{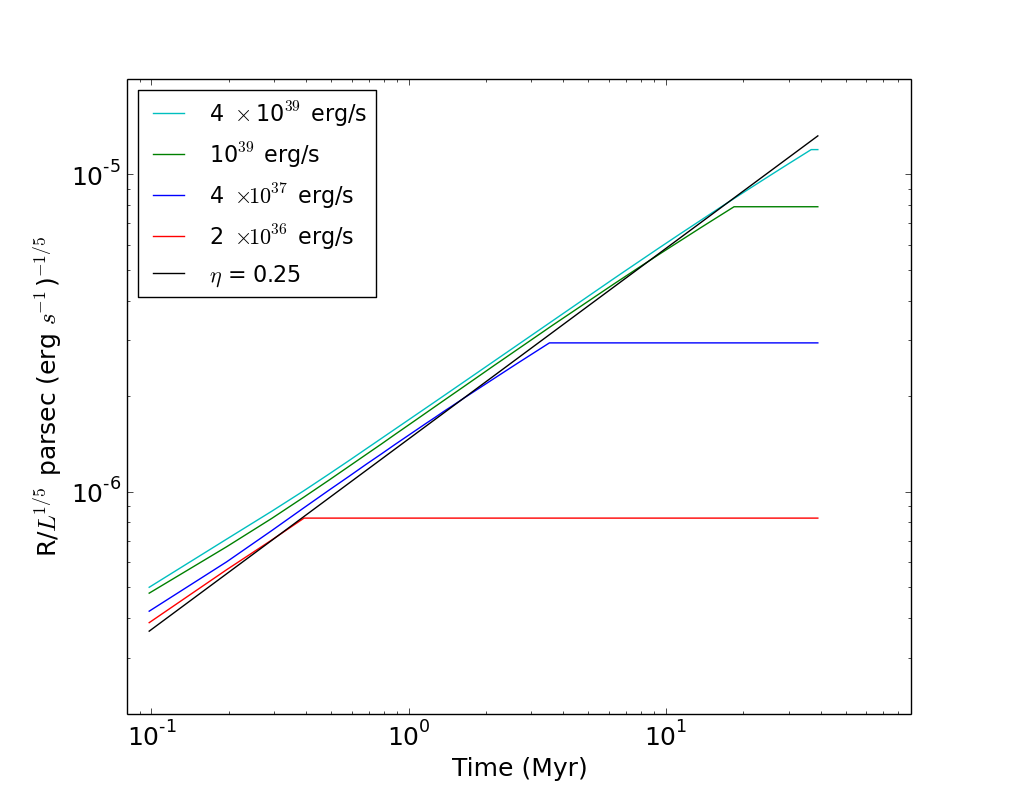}
\caption{Similar to Figure \ref{eta} except that this is for $\Delta r=0.1$ pc. }
\label{convergence2}
\end{figure}

This implies that $\eta$ has also reached convergence limit, and we show in Figure \ref{convergence2}
the logarithm of ratio of radius to $L^{1/5}$ vs.  time for different luminosities, superimposed on the expected evolution for $\eta=0.25$. The 
curves show (as in Figure \ref{eta}) that $\eta$ depends on luminosity rather weakly.

We should note that our runs do not simulate the effects of turbulent mixing in ISM, which might render 
numerical convergence ineffective (e.g., \citet{gentry2019, fierlinger2016}). However, those effects are beyond the scope of the present work.

\end{document}